\documentclass[12pt]{article}
\catcode`\@=11
\@addtoreset{equation}{section}

\global\arraycolsep=2pt
\usepackage{mathrsfs,amsbsy,amssymb,latexsym,amsfonts,amsmath,cite}
\usepackage{graphicx,color}
\usepackage{physics}
\usepackage{mathtools}
\usepackage{ulem}
\usepackage{caption}
\usepackage{here}

\usepackage{mathrsfs,amsbsy,amssymb,latexsym,amsfonts,amsmath,cite,bm}
\usepackage{graphicx,color}
\usepackage{mathtools}
\usepackage{enumerate}

\DeclareFontFamily{U}{BOONDOX-calo}{\skewchar\font=45 }
\DeclareFontShape{U}{BOONDOX-calo}{m}{n}{
  <-> s*[1.05] BOONDOX-r-calo}{}
\DeclareFontShape{U}{BOONDOX-calo}{b}{n}{
  <-> s*[1.05] BOONDOX-b-calo}{}
\DeclareMathAlphabet{\mathcalboondox}{U}{BOONDOX-calo}{m}{n}
\SetMathAlphabet{\mathcalboondox}{bold}{U}{BOONDOX-calo}{b}{n}
\DeclareMathAlphabet{\mathbcalboondox}{U}{BOONDOX-calo}{b}{n}

\newcommand{\red}[1]{#1}

\allowdisplaybreaks

\begin{document}

\renewcommand{\thefootnote}{\fnsymbol{footnote}}

\begin{flushright}
RIKEN-iTHEMS-Report-24, \\ 
KUNS-3023, \\ 
STUPP-24-273
\end{flushright}
\vspace*{0.5cm}

\begin{center}
{\Large \bf  Scaling law for membrane lifetime
}
\vspace*{1.5cm} \\
{\large  Osamu Fukushima$^{\sharp}$\footnote{E-mail:~osamu.fukushima@riken.jp},
Tomohiro Shigemura$^{\flat}$\footnote{E-mail:~shigemura@gauge.scphys.kyoto-u.ac.jp},
and Kentaroh Yoshida$^{\natural}$\footnote{E-mail:~kenyoshida@mail.saitama-u.ac.jp
}} 
\end{center}

\vspace*{0.4cm}

\begin{center}
$^{\sharp}${\it iTHEMS, RIKEN, Wako, Saitama 351-0198, Japan}
\end{center}
\begin{center}
$^{\flat}${\it Department of Physics, Kyoto University, Kyoto 606-8502, Japan}
\end{center}
\begin{center}
$^{\natural}${\it Graduate School of Science and Engineering,\\Saitama University, 255 Shimo-Okubo, Sakura-ku, Saitama 338-8570, Japan}
\end{center}

\vspace{1cm}

\begin{abstract}
Membrane configurations in the Banks-Fischler-Shenker-Susskind matrix model are unstable due to the existence of flat directions in the potential and the decay process can be seen as a realization of chaotic scattering. In this note, we compute the lifetime of a membrane in a reduced model. The resulting lifetime exhibits scaling laws with respect to energy, coupling constant and a cut-off scale. We numerically evaluate the scaling exponents, which cannot be fixed by the dimensional analysis. Finally, some applications of the results are discussed.
\end{abstract}

\setcounter{footnote}{0}
\setcounter{page}{0}
\thispagestyle{empty}

\newpage

\tableofcontents

\renewcommand\thefootnote{\arabic{footnote}}

\section{Introduction}

A promising candidate of non-perturbative formulation of superstring is the Banks-Fischler-Shenker-Susskind (BFSS) matrix model \cite{Banks:1996vh}. This is a one-dimensional matrix quantum mechanics and can also be obtained as a matrix regularization of the supermembrane theory \cite{dWHN}. It is known that this matrix model has membrane solutions \cite{dWHN} which are unstable due to the existence of flat directions in the potential \cite{dWLN}. The BFSS matrix model is chaotic in the sense that classical D0-brane dynamics exhibits non-vanishing Lyapunov exponents \cite{Arefeva:1997oyf}\footnote{There is a kind of massive deformation of the BFSS matrix model, which is called the Berenstein-Maldacena-Nastase (BMN) matrix model \cite{Berenstein:2002jq}. This model also exhibits classical chaos \cite{Asano:2015eha}.}.

\medskip 

In the preceding work \cite{Fukushima:2022lsd}, it has been shown that the membrane instability can be seen as chaotic scattering, which is a realization of transient chaos. For works for chaotic scattering in the string-theory context, see \cite{string1,string2,string3}. In particular, the initial configuration of a membrane is put inside the potential to describe the membrane decay so that the chaotic motion is included in the decay process. Hence this process should be referred to as the chaotic membrane decay. In \cite{Fukushima:2022lsd}, fractal structures in the time delay functions and the associated fractal dimensions have been computed numerically. These are characteristics of chaotic scattering \cite{review1,review2}. 

\medskip 

In this note, we will extend the previous analysis to examine membrane lifetimes. In particular, we show the scaling behavior of lifetimes with respect to energy, coupling constant and a cut-off scale, and then determine the scaling exponents. The resulting numerical values are non-trivial because they cannot be determined by the dimensional analysis. Also, the computation of the lifetime does not depend on particular initial conditions and its value is universal in contrast to the fractal dimensions computed in \cite{Fukushima:2022lsd}.  

\medskip 

This note is organized as follows. Section 2 provides a computational method to calculate the lifetime in chaotic scattering with a simple model, the four-hill model. In particular, we numerically clarify the lifetime scaling law. In section 3, we apply the method introduced in section 2 to a reduced model obtained from the BFSS matrix model. This model captures the dynamics of a simple membrane configuration and its lifetime can also be computed by introducing a cut-off for the membrane decay. Similarly, we show the lifetime scaling law numerically. Section 4 is devoted to conclusion and discussion.

\section{Lifetime scaling law in the four-hill model}

In this section, we investigate the chaotic decay in the four-hill model as a toy model and evaluate its lifetime. In particular, this lifetime depends on the system parameters and exhibits scaling laws in terms of them. We then numerically determine the scaling exponents.  

\subsection{The four-hill model}\label{sec:four-hill_potential_model}

\begin{figure}
    \centering
    \includegraphics[width=0.5\linewidth]{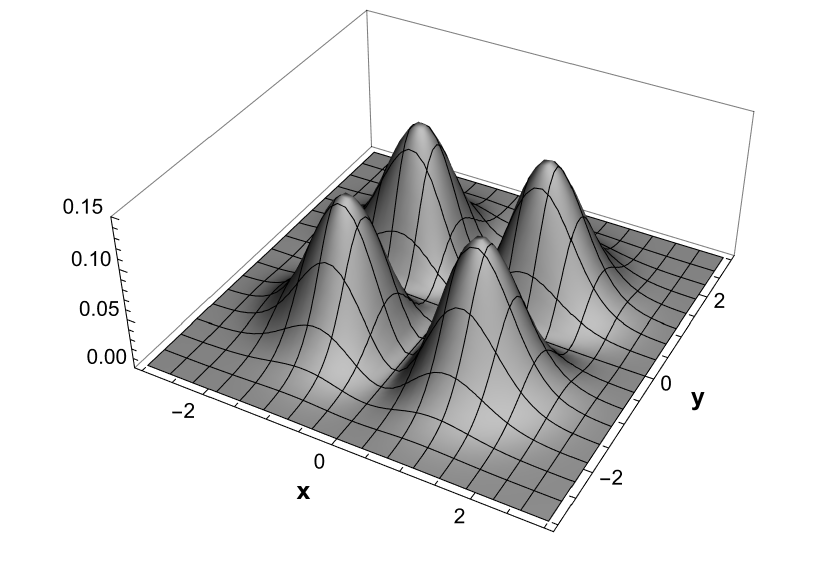}
\caption{\footnotesize The shape of the potential with $g=1$ and $ \sigma=1$\,.}\label{fig:four-hill_potential}
\end{figure}

Let us first introduce the four-hill model. This model describes the dynamics of a point particle moving on a $(x,y)$-plane with a certain potential. The Hamiltonian is given by 
\begin{align}
    H=\frac{p_x^2+p_y^2}{2m} + \frac{g}{2}\, x^2y^2\exp(-\frac{x^2+y^2}{\sigma^2})\,.
    \label{eq:four hill}
\end{align}
Here $p_x$ and $p_y$ are the canonical momenta, and $m$ and $g$ are a mass and coupling constant, respectively. For the positivity of the potential, $g$ is assumed to be positive. The parameter $\sigma$ is also a positive constant and measures the size of the chaotic region. The shape of the potential is depicted in Fig.\,\ref{fig:four-hill_potential}, where there exist four pillars and so this model is called the four-hill model. Note that this potential has the asymptotically flat region far apart from the pillars. 

\medskip 

The existence of the pillars is crucial to generate chaotic motions like in stadium billiards. The region which is relevant to chaos is called the chaotic region. This region is defined by the value of $\sigma$ like 
\begin{equation}
    -\sigma\leq x \leq \sigma\,, \qquad -\sigma \leq y \leq \sigma\,.  
    \label{region}
\end{equation}
With initial conditions set within the chaotic region, the particle motion is chaotic at first. Then after a while the particle escapes from the chaotic region to the asymptotically flat region. This process can be seen as a kind of particle decay including chaotic property. So this is called {\it chaotic decay} below. The fractal structure in the time delay functions and the associated fractal dimension are computed in \cite{review1,Fukushima:2022lsd}.  

\subsection{Lifetime}

In this note, we are interested in computing the lifetime of a particle associated with the chaotic decay. First of all, let us introduce the notion of {\it escape time} from the potential. Suppose that the initial conditions are taken inside the chaotic region (\ref{region}). Then the particle starts to move and exhibits chaos. After a while, it escapes from the chaotic region to the outside. The escape time refers to the time it takes for the particle to escape from the potential. 

\medskip 

\begin{figure}
~~~
    \begin{minipage}[ht]{0.4\linewidth}
    \centering
    \includegraphics[width=\linewidth]{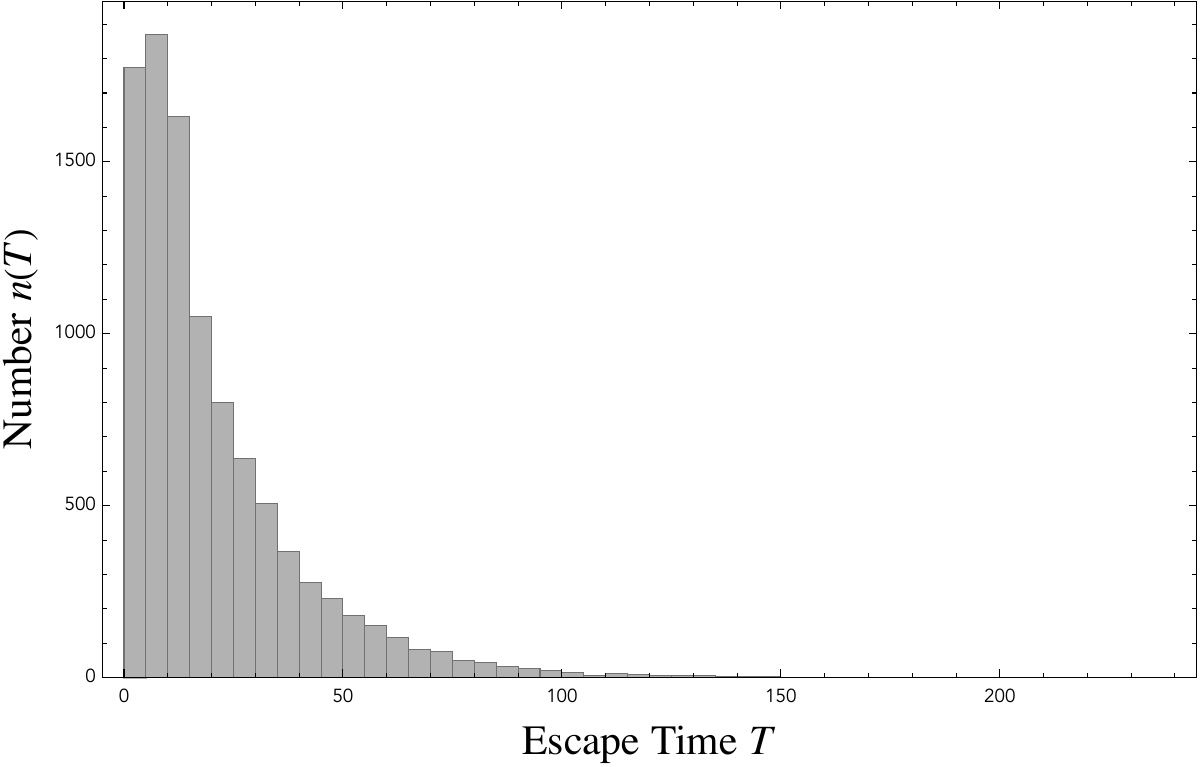}
    \end{minipage}
    \qquad 
   ~~~~ \begin{minipage}[ht]{0.4\linewidth}
    \centering
    \includegraphics[width=\linewidth]{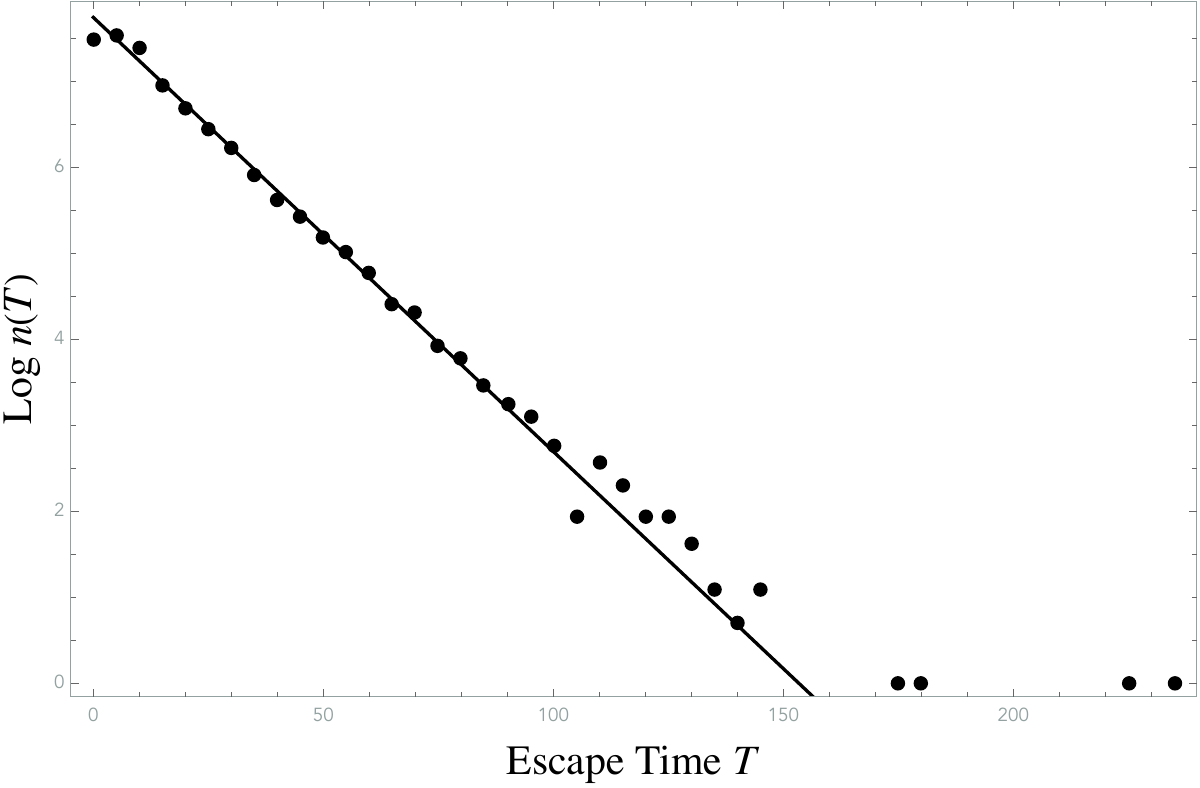}
    \end{minipage}
\caption{\footnotesize A histogram for event number $n(T)$ vs.\ escape time $T$ is drawn (left). The log of event number is fitted by a linear function (right). The parameter values are $E=0.1\,,~g=1\,,~\sigma=1$ and $m=1$\,.}
\label{fig:escape_time}
\end{figure}

An escape time for a single event can be computed by taking an initial condition inside the chaotic region (\ref{region}). By changing the initial condition and repeating this computation, we will get a list of escape times. By randomly choosing $N$ points of the initial conditions from the chaotic region (\ref{region}), a histogram can be drawn as in Fig.\,\ref{fig:escape_time} (left). Note that this histogram has been computed by setting $E=0.1\,,~m=1\,,~g=1\,,~\sigma=1$ and the total event number $N=10^4$\,. The horizontal axis is the escape time $T$ and the vertical axis is the number of events $n(T)$\,. 

\medskip 

Next, the histogram can be fitted by using the exponential form like 
\begin{eqnarray}
    n(T) \propto \exp\left(-\frac{T}{\tau}\right)\,, 
    \label{tau}
\end{eqnarray}
where $\tau$ is a real, positive constant and called the {\it lifetime} of a particle. This lifetime $\tau$ can be computed by the graph where the vertical axis is the log of the event number as in Fig.\,\ref{fig:escape_time} (right). The lifetime is obtained by fitting the points with a linear function and finding its gradient. As a result, the lifetime for $E=0.1\,,~g=1\,,~\sigma=1$ and $m=1$ is given by 
\begin{equation}
    \tau = 23.1 \pm 0.1 \,. \label{life}
\end{equation}

\medskip

It should be remarked that in Fig.\,\ref{fig:escape_time} (right)\,, the region with the escape time $ T <10$ does not obey the power law. This is because this short-time behavior comes from the fact that no classical chaos occurred. These non-chaotic processes have been ignored in the fitting. In Fig.\,\ref{fig:escape_time} (right) we also ignore the region with \red{$T>150$} in the fitting because the sample number in this region is too little.

\medskip 

The lifetime in (\ref{life}) has been computed for certain values of the parameters $E$\,, $g$\,, $l$ and $m$ and as a matter of course, the value of $\tau$ changes depending on these parameters. The next task is to study the parameter dependence of the lifetime numerically.

\subsection{Dimensional analysis}

Before starting numerical studies, let us make the dimensional analysis for the lifetime. To this end, we introduce three types of dimensions, i.e., length $[L]$\,, time $[T]$\,, and mass $[M]$\,, and then we find that the Hamiltonian (\ref{eq:four hill}) contains three dimensionful constants, mass $m$\,, coupling constant $g$ and the size of the chaotic region $\sigma$\,. 
For the dimensional analysis, in addition to them, we will use the energy $E$ of the system. 

\medskip 

The lifetime $\tau$ has dimension of time $[T]$\,. The dimension of $g$ is $\left[\frac{M L^2}{T^2} \cdot \frac{1}{L^4} = \frac{M}{L^2T^2}\right]$\,. The dimensions of $m$ and $\sigma$ are $[M]$ and $[L]$\,, respectively. Thus, if the lifetime is written as 
\begin{eqnarray}
\tau &\sim& m^a\, g^b\, E^c\, \sigma^d = \left[M^a \left(\frac{M}{L^2T^2}\right)^b \left(\frac{ML^2}{T^2}\right)^c L^d \right] 
\label{para} \\ 
&=& \left[
M^{a+b+c}\,L^{-2b+2c+d}\, T^{-2b -2c} \nonumber 
\right]
\,,
\end{eqnarray}
then the following three conditions should be satisfied for the four constants $a,b,c,d$~: 
\[
a+ b+ c = 0\,, \qquad -2b+2c +d = 0\,, \qquad -2b-2c = 1\,.
\]
There remains a single parameter. When $c$ is left as an undetermined constant, we obtain the following relations: 
\begin{equation}
a = \frac{1}{2}\,, \qquad b= -\frac{1}{2} -c\,, \qquad d = -1-4c\,. 
\label{constraint}
\end{equation}
As a result, the lifetime $\tau$ is evaluated as 
\begin{equation}
    \tau \sim \sqrt{\frac{m}{\sigma^2g}}\,\left(\frac{E}{\sigma^4 g}\right)^c\,. 
    \label{function1}
\end{equation}
where $c$ is the dimensionless constant. In general, the lifetime is expressed as
\begin{eqnarray}
    \tau 
    &=& \sqrt{\frac{m}{\sigma^2g}}\,f\left(\frac{E}{\sigma^4 g}\right) 
    \label{eq:lifetime_parameter_dependence_four-hill}
\end{eqnarray}
by using an arbitrary function $f(x)$\,. The next task is to find out the form of $f(x)$ by examining the energy-dependence of the lifetime $\tau$\,.

\subsection{Scaling law}

\begin{figure}[tbp]
    ~~~~\begin{minipage}[ht]{0.4\linewidth}
    \centering
    \includegraphics[width=1\linewidth]{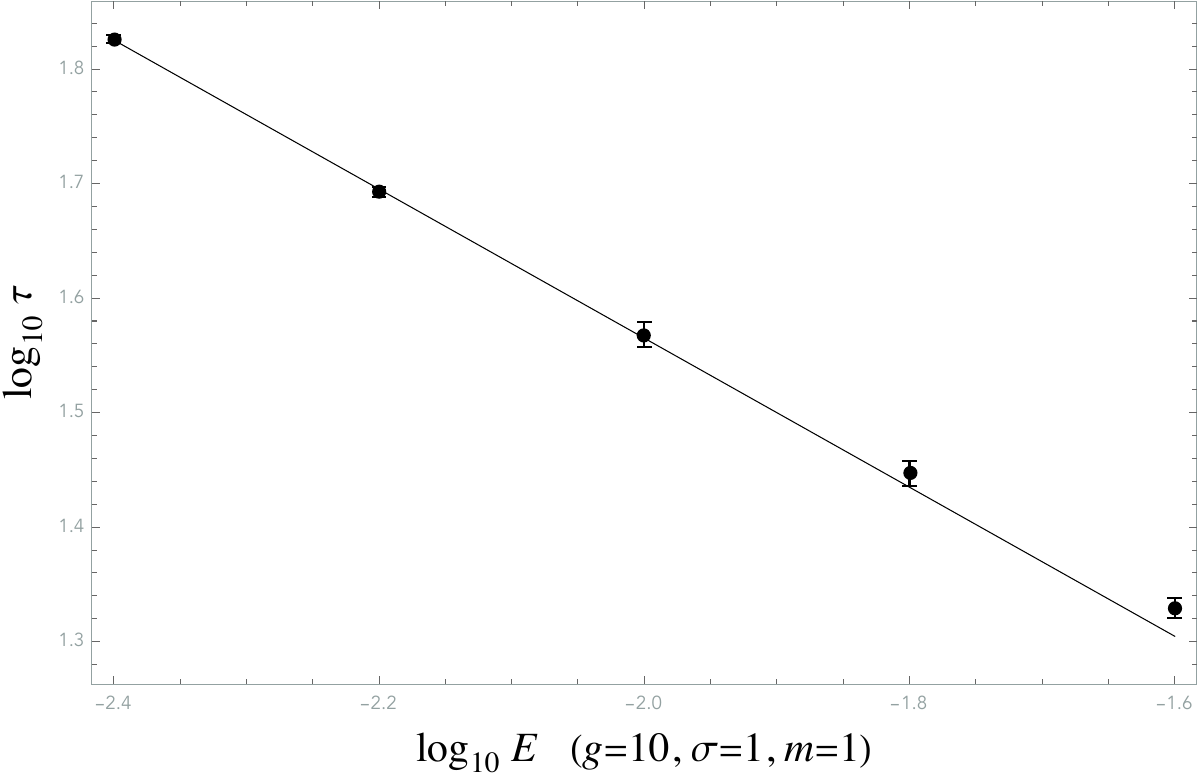} \\ 
    {\footnotesize (a) \quad  $E$-dependence } 
    \vspace*{0.5cm}\\
    \end{minipage}
    \qquad 
    ~~~~\begin{minipage}[ht]{0.4\linewidth}
    \centering
    \includegraphics[width=1\linewidth]{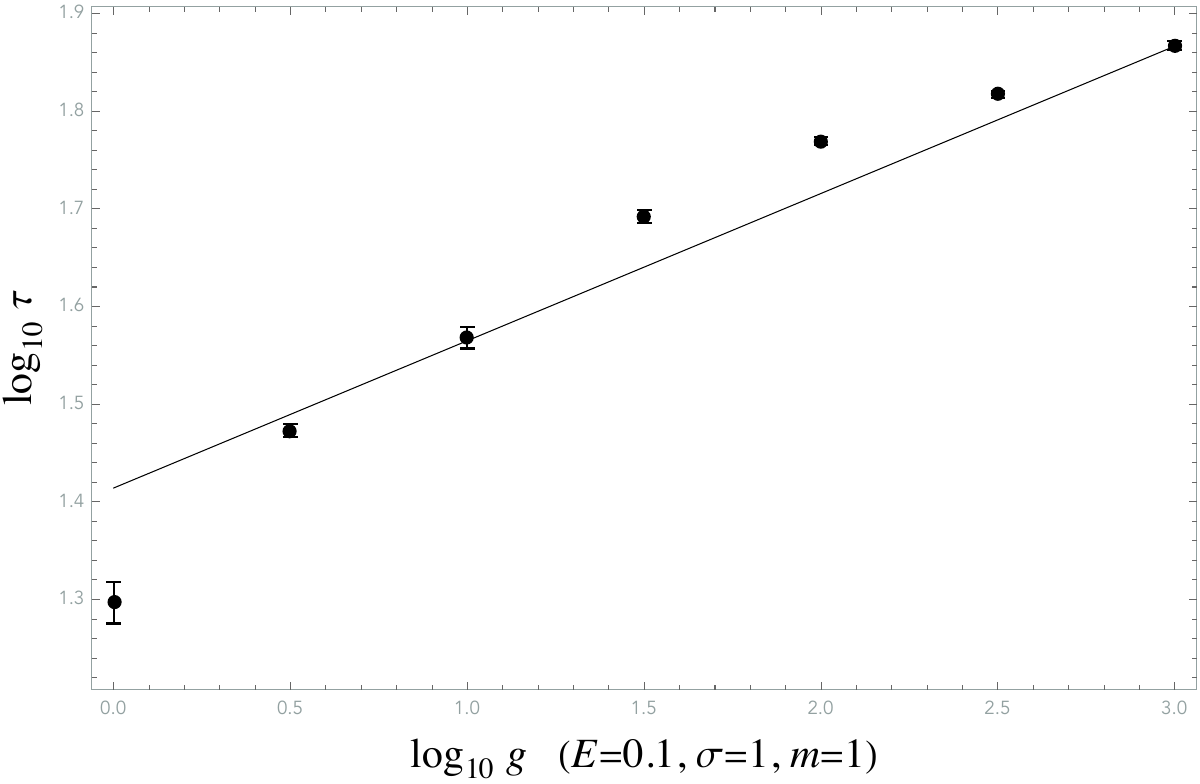} \\ 
    {\footnotesize (b) \quad  $g$-dependence } 
    \vspace*{0.5cm}\\
    \end{minipage}

    ~~~~\begin{minipage}[ht]{0.4\linewidth}
    \centering
    \includegraphics[width=1\linewidth]{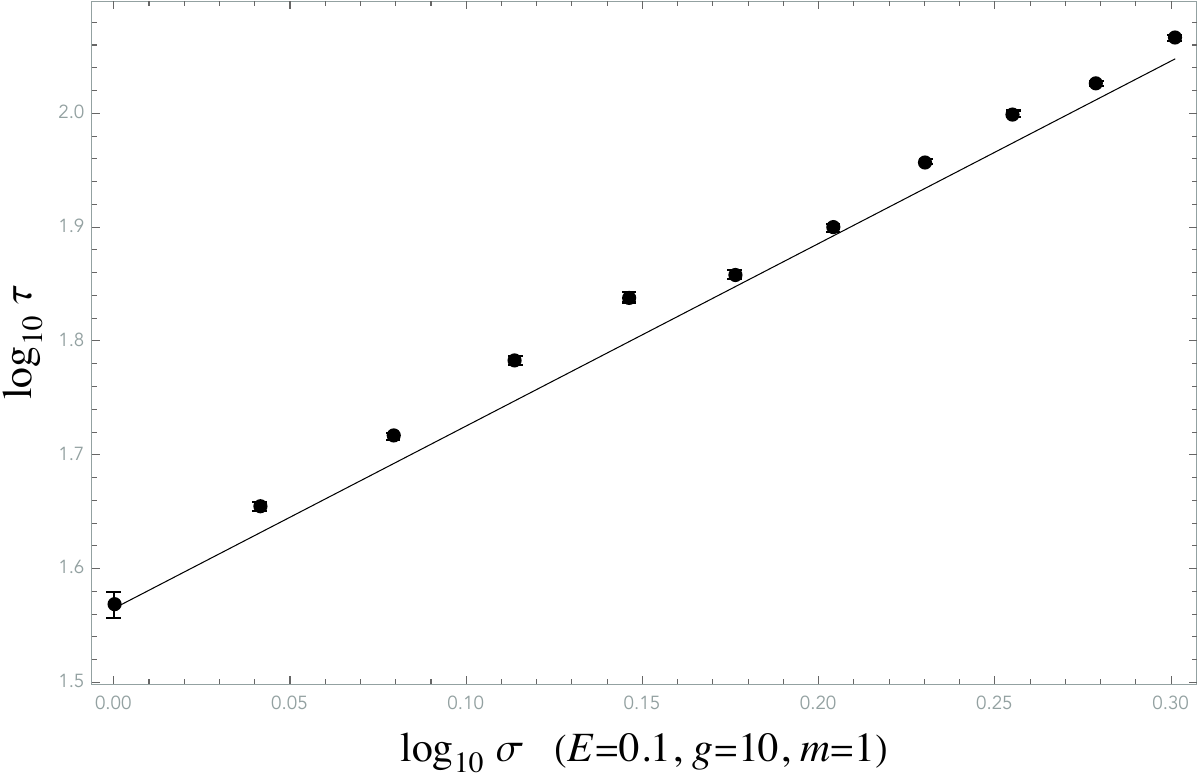} \\ 
    {\footnotesize (c) \quad $\sigma$-dependence } 
    \vspace*{0.5cm}\\
    \end{minipage}
    \qquad 
    ~~~~\begin{minipage}[ht]{0.4\linewidth}
    \centering
   \includegraphics[width=1\linewidth]{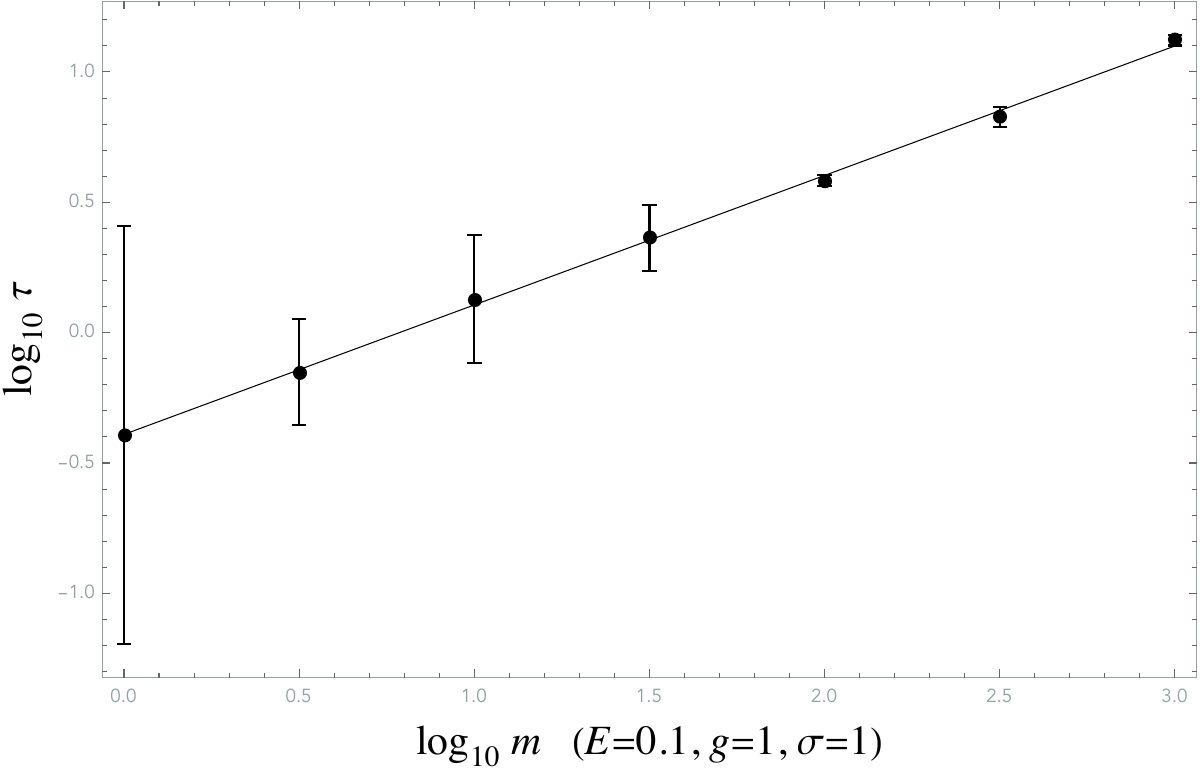} \\ 
    {\footnotesize (d) \quad $m$-dependence } 
    \vspace*{0.5cm}\\
    \end{minipage}
    \vspace*{-10pt}
\caption{\footnotesize Log-log plots of lifetime $\tau$ against one of $\{ E, g, \sigma, m \}$ with the other parameters fixed.}\label{fig:lifetime_four-hill_log_log_plots}
\end{figure}

Let us here investigate the form of the function $f(x)$ in (\ref{eq:lifetime_parameter_dependence_four-hill})\,. For that purpose, we will compute the lifetime as a function $\tau (E,g,\sigma,m)$ for every combination of following values of the parameters:
\begin{align}
    E &= \{10^{-2}\, , ~ 10^{-1.5}\, , ~ 10^{-1}\, , ~ 10^{-0.5}\, , ~ 1\}\,, 
    \\
    g &= \{1\, , ~ 10^{0.5}\, , ~ 10\, , ~ 10^{1.5}\, , ~ 10^{2}\, , ~ 10^{2.5}\, , ~ 10^{3}\}\,, 
    \\
    \sigma &= \{1\, , ~ 1.1\, , ~ 1.2\, , ~ 1.3\, , ~ 1.4\, , ~ 1.5\, , ~ 1.6\, , ~ 1.7\, , ~ 1.8\, , ~ 1.9\, , ~ 2\}\,, 
    \\
    m &= \{1\}\,.
\end{align}
The total number of the events is set to $N=10^4$ again. A numerical result for the $E$-dependence with $g=10$ and $\sigma=1$ is presented in Fig.\,\ref{fig:lifetime_four-hill_log_log_plots} (a). Similarly, results for $g$ and $\sigma$ are shown in Fig.\,\ref{fig:lifetime_four-hill_log_log_plots} (b) and (c)\,, respectively. All of these plots are fitted by using linear functions. These results indicate that the lifetime behaves simply as in (\ref{function1}) and the function $f(x) \propto x^c$ rather a non-trivial function. Therefore, by fitting all of the data with the form of the function (\ref{function1})\,, the value of $c$ is determined as 
\begin{eqnarray}
c=-0.651 \pm 0.032 \,.
\end{eqnarray}

Also, we have investigated the mass dependence for $E=0.1$\,, $g=1$\,, and $\sigma=1$ by choosing the values of $m$  from $m=\{1, 10^{0.5}, 10, 10^{1.5}, 10^2, 10^{2.5}, 10^3\}$ with $N=10^5$ initial conditions. This is a consistency check because the mass dependence is completely determined by the dimensional analysis like $\tau\propto m^\frac{1}{2}$ (i.e., $a=1/2$)\,. As shown in Fig.\,\ref{fig:lifetime_four-hill_log_log_plots} (d)\,, the mass dependence is computed as 
\begin{equation}
  a=0.496 \pm 0.018 \,.  
\end{equation}
This result is in good agreement with the exact value $a=1/2$\,.

\section{Scaling law for membrane lifetime}

In this section, let us compute a membrane lifetime in the BFSS matrix model. For simplicity, we will consider a simple membrane configuration. For this purpose, it is helpful to reduce the BFSS matrix model by imposing some ansatz. Then, by applying the method used for the four-hill model to this reduced model, we can compute a membrane lifetime. We find a scaling law for the resulting lifetime again and the scaling exponents are computed numerically. 

\subsection{Reducing the BFSS matrix model}\label{sec:chaotic-instability} 

Let us introduce a reduced model obtained from the BFSS matrix model below.  

\medskip 

The BFSS matrix model is a $1+0$ dimensional $SU(N)$ gauge theory with adjoint matters\footnote{\red{The original gauge group is $U(N)$\,, but the $U(1)$ is decoupled. Therefore, we take the group as $SU(N)$\,.}}. The bosonic part of the classical action is given by
\begin{align}
    S &=\red{\frac{1}{l_s g_s}}\int\! dt\, \tr( \frac{1}{2}\sum_{r=1}^{9}(DX^{r})^2 + \frac{1}{4}\sum_{r,s=1}^{9} \left[ X^{r},X^{s}\right]^2 )\,,
    \label{eq:BFSS action}
    \\
    DX^{r} &:= \frac{d}{dt}X^{r} - i [A,X^{r}]\,.
\end{align}
Each of $X^r$'s is an $N \times N$ hermitian matrix \red{and the action has the $SO(9)$ global $R$-symmetry. The constants $l_s$ and $g_s$ are string length and string coupling, respectively.} The gauge field $A$ is also a hermitian matrix and non-dynamical. The variation with respect to $A$ leads to the Gauss law constraint: 
\begin{equation}
 0= \sum_{r=1}^9[X^r, DX^r]\,. \label{Gauss}
\end{equation}
The equations of motion for $X^r$'s are given by 
\begin{equation}
0 = \frac{d^2}{dt^2}X^r + \sum_{s=1}^9[X^s,[X^s,X^r]]\,, \label{eom}
\end{equation}
where we have taken the static gauge $A=0$\,. 

\medskip 

Hereafter, we will consider the $N = 2$ case for simplicity and furthermore impose the following ansatz for solutions to the equations of motion \eqref{eom}:
\begin{align}
    X^1(t) &=x(t)\,\frac{\sigma^1}{2}\,, \qquad 
    X^2(t) =y(t)\,\frac{\sigma^2}{2}\,, \qquad 
    X^3(t) =z(t)\, \frac{\sigma^3}{2}\,,
    \label{eq:EoM ansatz}
    \\ 
    X^{s}(t) &=0 \qquad (s=\red{4},\dots,9)\,, \nonumber 
\end{align}
where $\sigma^i~(i=1,2,3)$ are the standard Pauli matrices and $x(t)$\,, $y(t)$ and $z(t)$ are real-valued functions. Note that this ansatz is consistent to the Gauss law constraint (\ref{Gauss})\,. 

\medskip 

Then the equations of motion \eqref{eom} are written in terms of $x(t)$\,, $y(t)$ and $z(t)$ as
\begin{align}
    0=&\frac{d^2}{dt^2}x(t)+x(t)(y(t)^2 + z(t)^2)\,,\notag \\ 
    0=&\frac{d^2}{dt^2}y(t)+y(t)(z(t)^2 + x(t)^2)\,, \\
    0=&\frac{d^2}{dt^2}z(t)+z(t)(x(t)^2 + y(t)^2)\,. \nonumber
\end{align}
These equations describe the motion of a spherical membrane solution \cite{Kabat:1997im}. 

\begin{figure}
    \centering
    \includegraphics[width=0.5\linewidth]{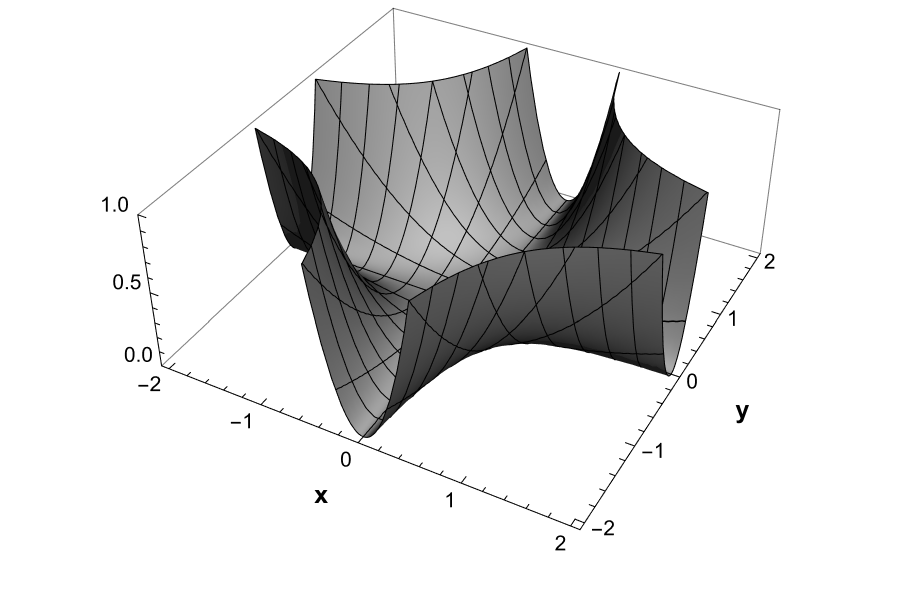}
\caption{\footnotesize The shape of the potential with $g=1$.}\label{fig:xy_potential}
\end{figure}

\medskip

In the following, by setting $z(t)=0$\,, we will focus upon the dynamics of $x(t)$ and $y(t)$\,. The resultant equations of motion can be derived from the Hamiltonian
\begin{align}
    H=\frac{p_x^2+p_y^2}{2m} + \frac{g}{2}\, x^2y^2\,.
    \label{eq:xy potential}
\end{align}
This model has already been discussed in some contexts \cite{Arefeva:1997oyf,Matinyan:1981dj}\footnote{For an argument on a similar reduced model, see \cite{Helling:2000kz}.}. Note that this Hamiltonian is obtained from the four-hill model (\ref{eq:four hill}) by dropping off the exponential factor. 

\medskip 

The potential in \eqref{eq:xy potential} has flat directions as depicted in Fig.\,\ref{fig:xy_potential} along which $|x|$ and $|y|$ can become infinitely large. This means that the diameter of the membrane in either direction can expand to any size within a finite time. Generally, the membrane is not expected to extend infinitely and is assumed to split into two pieces at a characteristic length scale $l$, which we refer to as the cut-off scale. This splitting process is regarded as a decay process hereafter as in \cite{Fukushima:2022lsd}. 

\medskip 

To study the property of the membrane instability, we consider time evolution until $x$ and $y$ reach the region with 
\begin{equation}
    \sqrt{x^2+y^2}\, > l\,.  
    \label{decay}
\end{equation}
We shall refer to this region (\ref{decay}) as the decay region. Once that region is reached, the membrane is considered to have decayed. The time the membrane takes to decay is referred to as {\it decay time} $\tilde{T}$, which is analogous to the escape time $T$ in the four-hill model.

\subsection{Lifetime}

By identifying the decay process as described in the previous subsection, we can consider the associated lifetime. By computing decay times for the set of initial conditions and making a histogram, the lifetime can be evaluated as in the four-hill case. However, in the first place, we need to determine the chaotic region in the model (\ref{eq:xy potential}). The decay region (\ref{decay}) is much bigger than the chaotic region (\ref{region}) because it contains non-chaotic region. In the model (\ref{eq:xy potential})\,, there is no dimensionful parameter measuring the size of the chaotic region such as $\sigma$ in the four-hill case. Our idea here is to define the chaotic region in the model (\ref{eq:xy potential}) by using $E/g$ (instead of $\sigma$) like 
\begin{equation}
 |x|\,,~|y| < \left(\frac{E}{g}\right)^{\frac{1}{4}} \label{region2}
\end{equation}
based on the potential shape. Note here that the decay time $\tilde{T}$ is measured once the relation (\ref{decay}) is realized. This condition is in contrast to the four-hill model where the escape time $T$ is measured when a particle escaped from the chaotic region \red{(\ref{region2})}.

\medskip

\begin{figure}
~~~
    \begin{minipage}[ht]{0.4\linewidth}
    \centering
    \includegraphics[width=\linewidth] {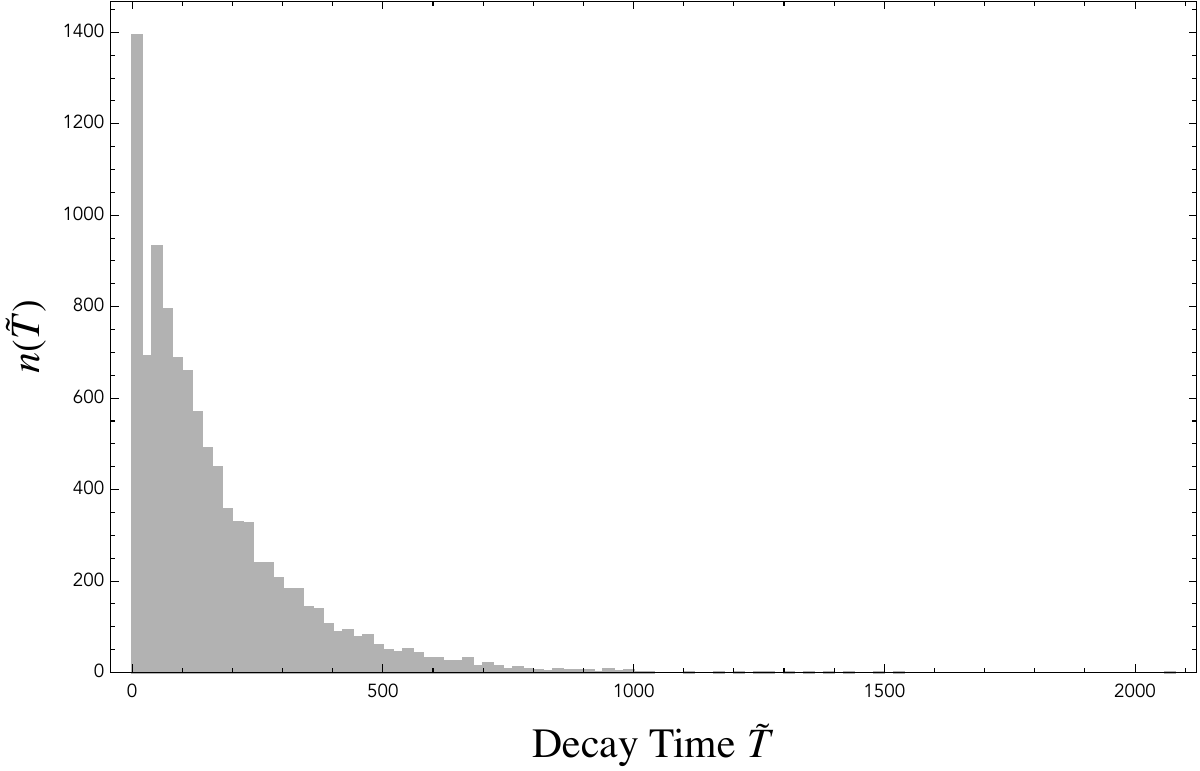}
    \end{minipage}
    \qquad 
    ~~~~\begin{minipage}[ht]{0.4\linewidth}
    \centering
    \includegraphics[width=\linewidth]{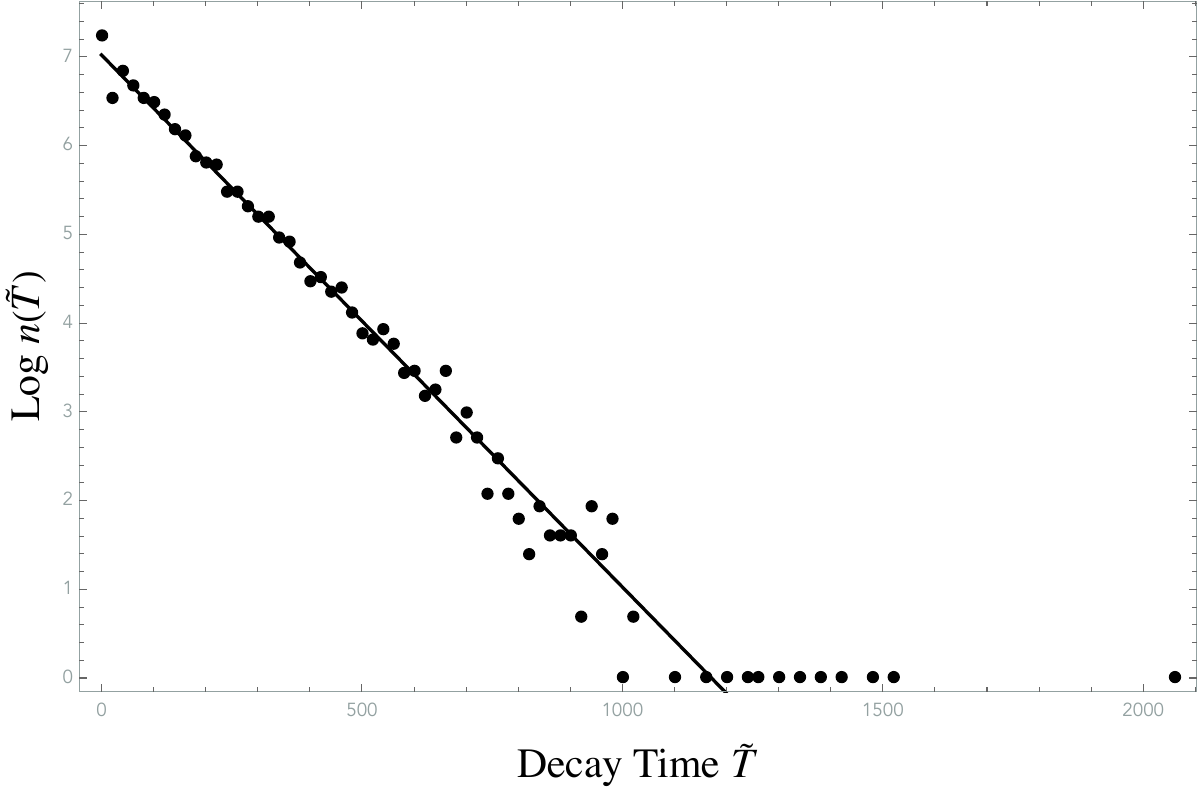}
    \end{minipage}
\caption{\footnotesize A histogram for event number $n(\tilde{T})$ vs.\ decay time $\tilde{T}$ is drawn (left). The log of event number is fitted by a linear function (right). The parameter values are $E=0.1\,,~g=1\,,~l=5$ and $m=1$\,.}
\label{fig:decay_time}
\end{figure}

Then $N=10^4$ initial conditions are taken from the region \red{(\ref{region2})}\,. The resulting histogram is presented in Fig.\,\ref{fig:decay_time} (left)\,. The log of the event number can be fitted by using a linear function again. According to the definition (\ref{tau})\,, the lifetime $\tau$ is evaluated as 
\begin{eqnarray}
    \tau = 166.8 \pm 0.1 \,,
\end{eqnarray}
\red{for $E=0.1$\,, $g=1$\,, $l=5$ and $m=1$\,.} The next task is to study the parameter dependence of the lifetime again.

\subsection{Dimensional analysis}

The dimensional analysis for the membrane lifetime is quite similar to the four-hill case. In the Hamiltonian (\ref{eq:xy potential}) contains two dimensionful constants, mass $m$ and coupling constant $g$\,. For the dimensional analysis, in addition to them, we will use the energy $E$ of the system and the cut-off scale $l$ as the parameters to characterize the system scale. 

\medskip 

The lifetime is evaluated by using $l$\,, instead of $\sigma$ in the four-hill case, like\footnote{For the dimensional analysis for the Lyapunov exponent, see \cite{Hashimoto:2021afd}. }  
\begin{eqnarray}
    \tau 
    &=& \sqrt{\frac{m}{l^2g}}\,h\left(\frac{E}{l^4 g}\right) 
    \label{eq:lifetime_parameter_dependence_xy}
\end{eqnarray}
by using an arbitrary function $h(x)$\,. We will try to find out the form of $h(x)$ by examining the energy-dependence of the lifetime $\tau$ again. This form is a model-dependent element, hence we may obtain a different result in comparison to the four-hill case.

\subsection{Scaling law}

We have numerically investigated the parameter-dependence of the lifetime $\tau (E,g,l,m)$\,. The values of the parameters $E$\,, $g$\,, $l$ and $m$ have been taken from the following:  
\begin{equation}
    \begin{split}
        E &= \{10^{-2}\,, ~10^{-1.5}\,, ~10^{-1}\,,~ 10^{-0.5}\,, ~1\}\,,
        \\
        g &= \{1\,, ~10^{0.5}\,,~ 10\,, ~10^{1.5}\,, ~10^{2}\,, ~10^{2.5}\}\,,
        \\
        l &= \{2.5\,,~3\,,~3.5\,,~4\,,~4.5\,,~5\,,~5.5\,,~6\,,~6.5\,,~7\,,~7.5\}\,,
        \\
        m &= \{1\}\,.
    \end{split}
    \label{eq:parameter_values_for_fitting_xy}
\end{equation}
Then $N=10^4$ initial conditions have been taken for each set of parameters.

\medskip

\begin{figure}[tbp]
   ~~~~ \begin{minipage}[ht]{0.4\linewidth}
    \centering
   \includegraphics[width=1\linewidth]{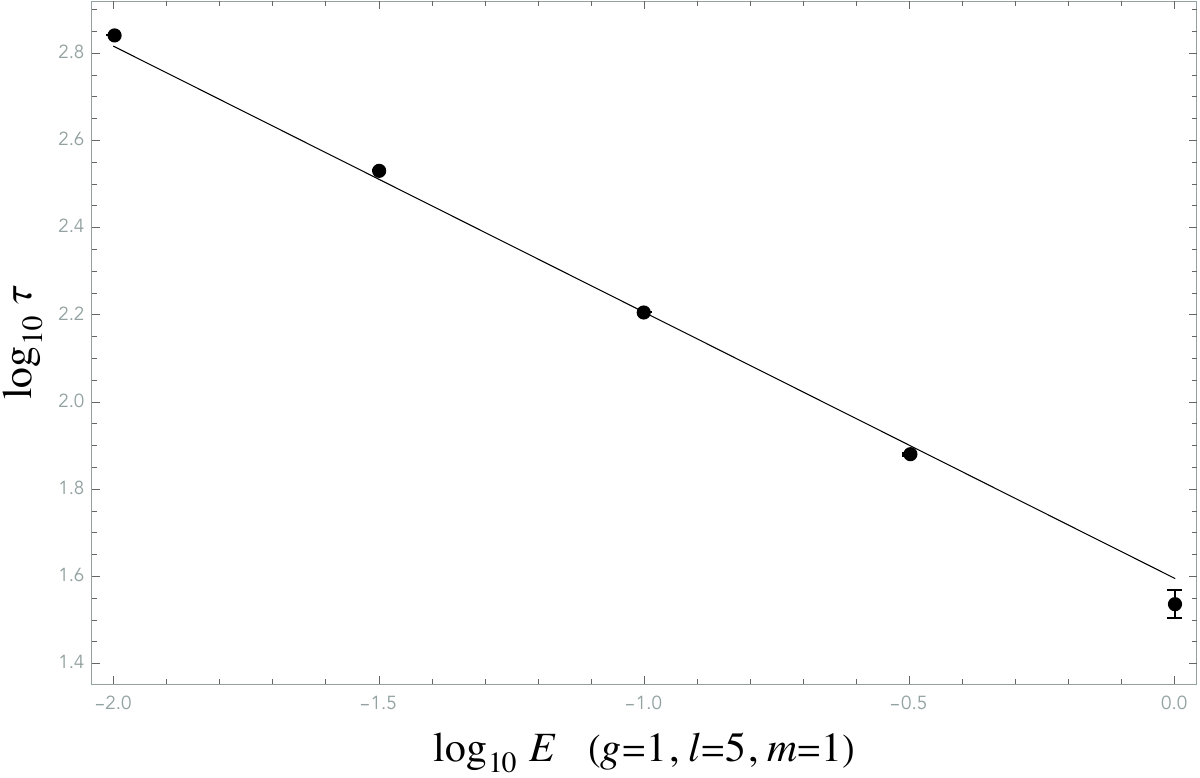} \\ 
    {\footnotesize (a) \quad  $E$-dependence } 
    \vspace*{0.5cm}\\
    \end{minipage}
    \qquad 
    ~~~~\begin{minipage}[ht]{0.4\linewidth}
    \centering
   \includegraphics[width=1\linewidth]{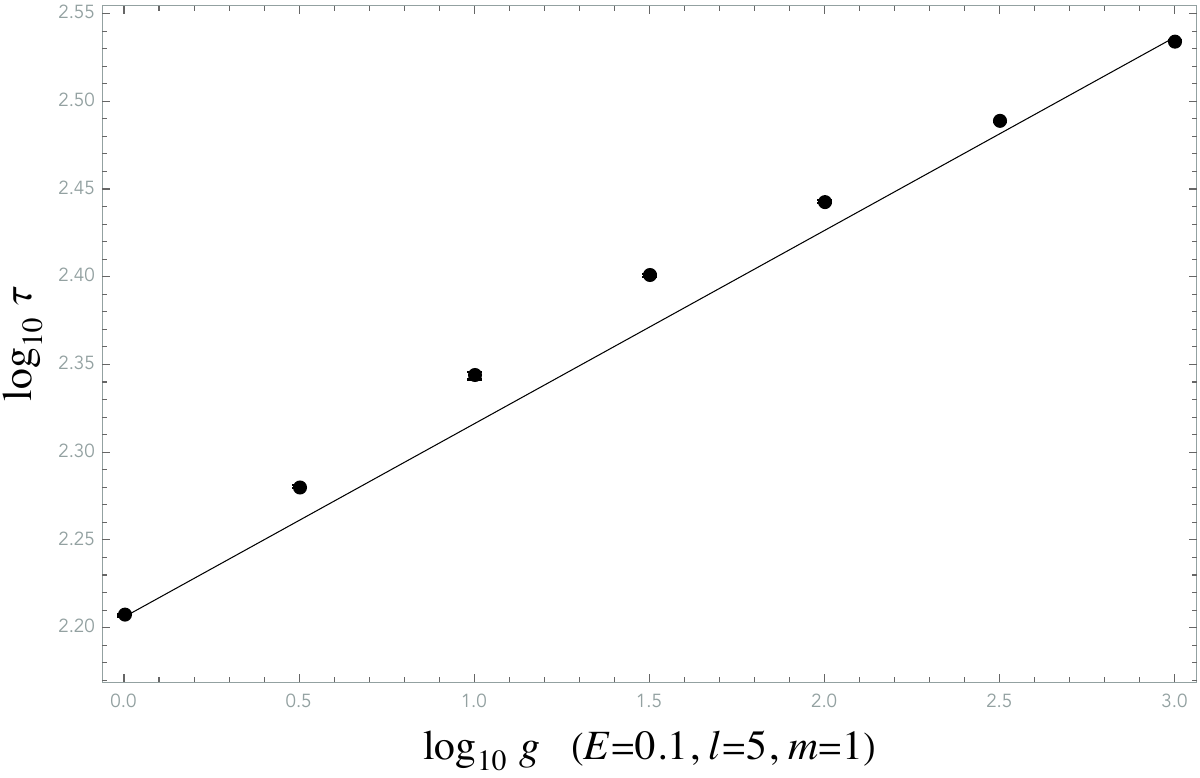} \\ 
    {\footnotesize (b) \quad  $g$-dependence } 
    \vspace*{0.5cm}\\
    \end{minipage}

   ~~~~ \begin{minipage}[ht]{0.4\linewidth}
    \centering
    \includegraphics[width=1\linewidth]{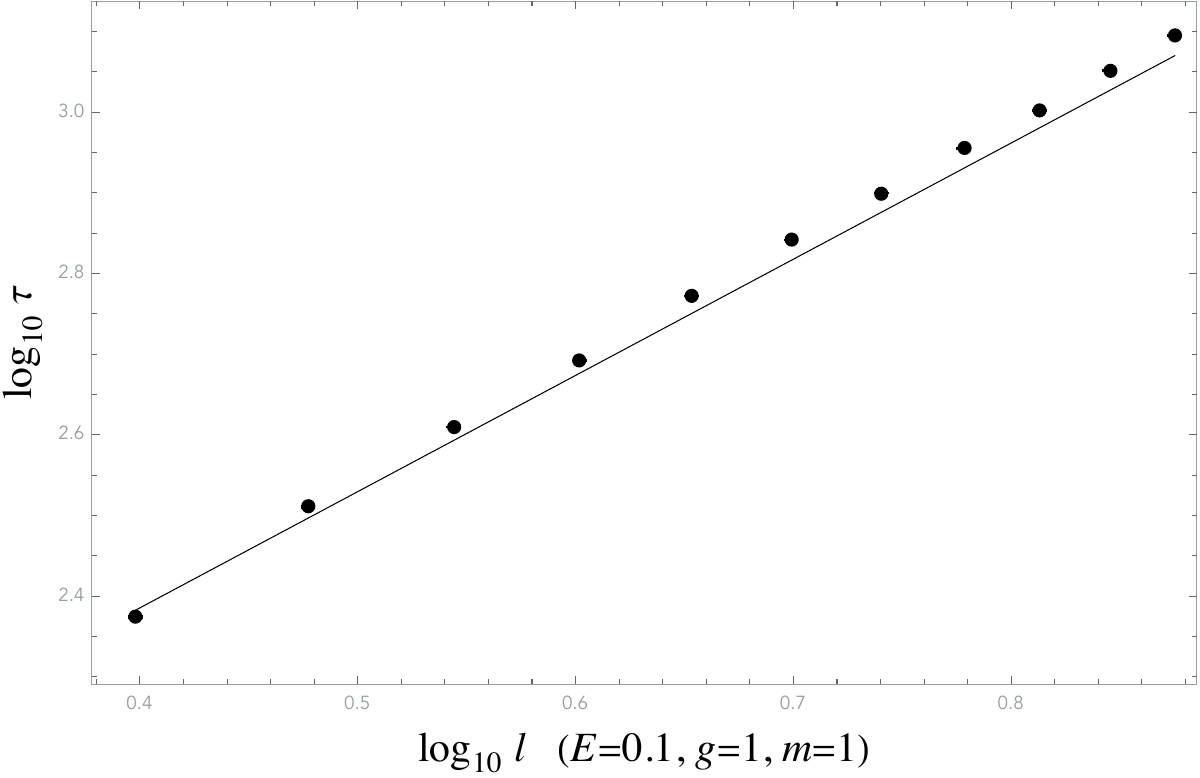} \\ 
    {\footnotesize (c) \quad $l$-dependence } 
    \vspace*{0.5cm}\\
    \end{minipage}
    \qquad 
   ~~~~\begin{minipage}[ht]{0.4\linewidth}
    \centering
     \includegraphics[width=1\linewidth]{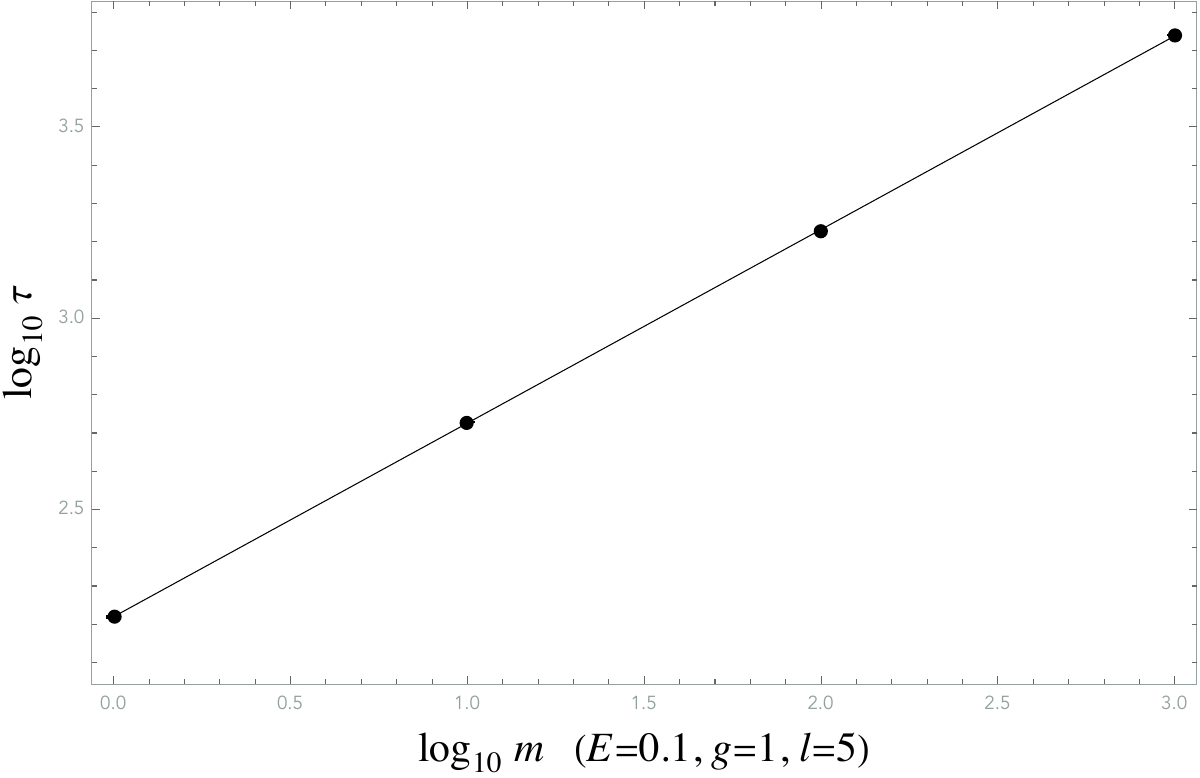} \\ 
    {\footnotesize (d) \quad $m$-dependence } 
    \vspace*{0.5cm}\\
    \end{minipage}
    \vspace{-10pt}
\caption{\footnotesize Log-log plots of lifetime $\tau$ against one of $\{ E, g, l, m \}$ with the other parameters fixed.}\label{fig:lifetime_xy_log_log_plots}
\end{figure}

The log-log plots of $\tau (E,g,l,m)$ against one of the parameters $\{ E, g, l, m \}$ with the others fixed are presented in Fig.\,\ref{fig:lifetime_xy_log_log_plots} (a)-(c). As a result, we have found that the lifetime can be expressed as a simple form,
\begin{equation}\label{eq:lifetime}
    \tau (E,g,l,m) \propto E^{\alpha} g^{-\alpha -\frac{1}{2}} m^{\frac{1}{2}} l^{-4\alpha -1}\,,
\end{equation}
as in the four-hill case. Then, by fitting all of the data, the value of $\alpha$ has been fixed as 
\begin{equation}
\alpha=-0.622 \pm 0.034 \,.     
\end{equation}
This is a non-trivial result which cannot be determined by the dimensional analysis. The undetermined function $h(x)$ in the dimensional analysis has been revealed as a simple form $h(x) \propto x^{\alpha}$\,. 

\medskip

Also, we have investigated the $m$-dependence with $E=0.1$\,, $g=1$ and $\sigma=1$ by changing $m$ like 
\begin{align}
m=\{1\,,~ 10\,,~ 10^2\,,~ 10^3\}
\end{align}
with $N=10^4$ initial conditions. Then the numerical result in Fig.\,\ref{fig:lifetime_xy_log_log_plots} (d) leads to 
\begin{equation}
   \tau \propto m^a\,, \qquad a = 0.506 \pm 0.003 \,.
\end{equation}
This result nicely agrees with the expected behavior $\tau\propto m^{\frac{1}{2}}$\,.

\newpage 

\section{Conclusion and Discussion}\label{sec:conclusion}

We have studied the lifetimes for membrane configurations in the BFSS matrix model by computing the histogram of the decay times. The results, in contrast to the Lyapunov exponents, do not depend on the  particular choice of initial conditions. In particular, dimensional analysis alone suggests that the lifetime could depend on an arbitrary function of the dimensionless energy. However, we have revealed that the function is just a power of the energy and the lifetime exhibits scaling laws with respect to energy, coupling constant and a cut-off parameter. The exponents have been determined numerically.

\medskip 

There are some future directions. A key challenge is to analytically prove the parameter dependence of lifetime. Since the scaling behavior is very special, there must be some fundamental reason behind it. This may be related to the membrane wave function \cite{Piljin}. As an application, it would be intriguing to use the membrane lifetime to study the real-time evaporation of matrix black holes as in \cite{Berenstein:2021gfx}.

\medskip 

It is also interesting to seek for a possible relation between the lifetime and the Lyapunov exponent associated with the classical chaos generated in the chaotic region. The Lyapounov exponent in the model (\ref{eq:xy potential}) also exhibits a peculiar energy-dependence \cite{Hashimoto:2021afd}\,. This behavior may be related to the scaling of lifetime. Studying the relation between the lifetime and other universal measures of chaotic scattering, such as the distribution function derived from scattering amplitudes~\cite{Bianchi:2022mhs}, is also an intriguing avenue of exploration.

\medskip

Finally, the fractal dimension computed in \cite{Fukushima:2022lsd} should be related to the lifetime. As discussed in \cite{saddle-dimension}, the fractal dimension and the lifetime are related in the ergodic system. Even in the Hamiltonian system (\ref{eq:xy potential})\,, which is not ergodic, there may be a connection in the form of inequality. We hope that we could report the result in the near future.

\subsection*{Acknowledgments}

K.\,Y.\ is very grateful to Gordon Semenoff for the kind hospitality during his stay at UBC, where discussions were very beneficial in completing this research.
The work of O.\,F.\ was supported by RIKEN Special Postdoctoral Researchers Program
and JSPS Grant-in-Aid for Research Activity Start-up No.\,24K22890.
The work of T.\,S.\ was supported by JST SPRING, Grant Number JPMJSP2110.
The work of K.\,Y.\ was supported by MEXT KAKENHI Grant-in-Aid for Transformative Research Areas A “Machine Learning Physics” No.\,22H05115, and JSPS Grant-in-Aid for Scientific Research (B) No.\,22H01217 \red{and (C) No.\,25K07313}.

\appendix

\section*{Appendix}

\section{\red{Parameters in the BFSS matrix model}}

Here, let us read off the relation among the parameters $m$ and $g$ in the reduced model (\ref{eq:xy potential}) and the string theoretical parameters, $l_s$ (string length) and $g_s$ (string coupling). 

\medskip 

The exact expression of the Lagrangian $L$ of the BFSS matrix model is given by 
\begin{eqnarray}
L = \frac{1}{l_s g_s}{\rm tr}\left[
\frac{1}{2} \sum_{r=1}^9(D X^r)^2 + \frac{1}{4 (2\pi l_s^2)^2} \sum_{r,s=1}^9[X^r,X^s]^2
\right]\,.  
\end{eqnarray}
Note here that the action (\ref{eq:BFSS action}) can be reproduced by setting $2\pi l_s^2 =1$\,. Then the Hamiltonian $H$ is given by 
\begin{eqnarray}
H = {\rm tr}\left[
\frac{l_s g_s}{2} \sum_{r=1}^9 (P_r)^2 - \frac{1}{16\pi^2l_s^5 g_s} \sum_{r,s=1}^9[X^r,X^s]^2
\right]\,, \label{H}
\end{eqnarray}
where $P_r~(r=1,\ldots,9)$ are the canonical momenta. 

\medskip 

By comparing the expression (\ref{H}) with the reduced Hamiltonian (\ref{eq:xy potential})\,, we can read off the following relations:
\begin{eqnarray}
m = \frac{1}{l_s g_s}\,, \qquad g = \frac{1}{4\pi^2 l_s^5g_s}\,. 
\end{eqnarray}
Then the lifetime is expressed as 
\begin{align}
\tau  
\propto&\,
E^{\alpha} \left(\frac{1}{ l_s^5g_s}\right)^{-\alpha -\frac{1}{2}} 
    \left(\frac{1}{l_s g_s}\right)^{\frac{1}{2}} 
    l^{-4\alpha -1}
    = (g_s E l_s)^{\alpha}  \left(\frac{l}{l_s}\right)^{-4\alpha-1} l_s\,.  
\end{align}
Note here that the cut-off $l$ remains as it is. Since our analysis is carried out at the classical level, $l$ is taken to be independent of $l_s$ and $g_s$.
At the quantum mechanical level, it would of course depend on these, but this issue is beyond our scope here.

\end{document}